\documentclass{elsart}

\usepackage{natbib}

\input psfig.sty

\begin{document}

\runauthor{McMahon et al}

\begin{frontmatter}
\title{The INT Wide Field Imaging Survey(WFS)}
\author[ioa]{R.G. McMahon}
\author[ing]{N.A. Walton,}
\author[ioa]{M.J. Irwin,}
\author[ioa]{J.R. Lewis,}
\author[ioa]{P.S. Bunclark,}
\author[ioa]{D.H.P. Jones,}
\author[ioa]{R.G. Sharp}

\address[ioa]{Institute of Astronomy, University of Cambridge, Cambridge, CB3 OHA, UK}
\address[ing]{Isaac Newton Group, LaPalma, Canary Islands, Spain}

\begin{abstract}
  The 2.5m Isaac Newton Telescope(INT) is currently being used to
  carry out a major multi-colour, multi-epoch, CCD based
  wide field survey over an area of $\sim$100deg$^2$. The survey
  parameters have been chosen to maximise scientific return
  over a wide range
  of scientific areas and to complement other surveys being carried
  out elsewhere.  Unique aspects of the survey is that it concentrates
  on regions of sky that are easily accessible from telescopes in both
  Northern and Southern terrestrial hemispheres and that it the first
  public survey to use filters similar to that being used by the Sloan
  Digital Sky Survey.  A major aim of the the INT Wide Field Survey
  program is to bridge the gap between the all-sky photographic 2 and 3
  band
  surveys such as the Palomar and UK Schmidt sky surveys and the
  ultra-deep keyhole surveys such as the Hubble Deep Field(Williams et al, 1996). Apart for
  the science that can be derived directly from the optical data, the
  datasets will provide ideal targets lists for multi-object followup
  with fibre and slit based systems(eg GMOS, 2DF, WYYFOS, FMOS)
  based systems on 4m and 8m  class telescopes.
\end{abstract}

\end{frontmatter}

\section{Introduction}

Astronomy is basically an observational science, and the development and 
advancement of the subject has relied heavily on surveys of the sky at 
optical wavelengths to expand our knowledge of the observable Universe.  
However despite the considerable advances in optical detector technology 
very little improvement has been made in large
scale surveys beyond those available in the
1950's when the Palomar Sky Survey was carried out. A photographic plate 
taken
on a 1.2-m Schmidt telescope is sky limited in about 1 hour
but is only 1-2\% efficient.  Thus our current best 
wide field optical sky surveys are equivalent to no more that a 
$\sim$60 second glance at
the Universe with modern CCDs using a 1m telescope. In spite of the
inherent limitations of photographic plates, they are still
used for major scientific programs. In recent years, this has
been primarily due to the availability of online digital atlas
images and catalogues based on these photographic material
eg http://www.ast.cam.ac.uk/\~{}apmcat, http://skyview.gsfc.nasa.gov/.

In an effort to rectify this apparently dismal situation, and to provide 
necessary underpinning imaging requirements for the  
8m telescope era, the 2.5m Isaac
Newton Telescope on LaPalma is being used to carry out a series
of wide field imaging programs under the generic title
of the INT Wide Field Survey(WFS) project.
The WFS project
consists of a series of independent survey programs with distinct aims
as we outline below.
The WFS project takes into account both surveys like SDSS
(Gunn \& Weinberg 1995) which also uses a 2.5m telescope and
has an exposure time of $\sim$60 seconds,
and other CCD based surveys
such as those that are being carried out by NOAO (http://www.noao.edu/) and 
ESO (http://www.eso.org/, Nonino etal 1999).
The unique
elements of the INT survey are: (i) optimal choice of fields so
that most are easily visible from telescopes in both hemispheres;
(ii) inclusion of U band; (iii) large area (iv) temporal information;
(iv) good overlapping coverage with existing deep radio surveys ie FIRST, WENSS;
(v) wide RA coverage optimised for efficient follow-up;
(vi) choice of SDSS bandpasses for longevity.

This article briefly describes the
Wide Field Survey (WFS) program. This is a peer reviewed survey program
that aims to provide deep high quality CCD data to the
community both quickly and in a convenient form.

\section{The INT Wide Field Survey}

The concept of the WFS originated in 1991 within the context
of the science case for a CCD mosaic for the Isaac Newton 
Telescope. Formal approval for the survey program began
with a proposal to the ING Board in October 1997.
The primary goal was to exploit the excellent capabilities of
a recently completed CCD based mosaic that effectively filled the
unvignetted focal plane of the 2.5m Isaac Newton Telescope
(see Figure~\ref{fig_wfc_geometry}).
The immediate aim was to carry out a major CCD based multi-colour survey
in a timely fashion over a period of 4--5 years and allow
instant and easy access to the processed data to facilitate its
rapid scientific
exploitation

The WFS proposal was approved by the ING Board in October 1997 with a
subsequent `Announcement of Opportunity' closing in March 1998. 
Conditions of solicitation included that the survey data 
is available to all UK and NL based astronomers in near
real-time. Raw data is typically available as taken, whilst the
pipeline reduced data is available after one month. Subsequently the
raw and processed data is
available to the rest of the astronomical community after one year. 
Pipeline processing of the data is the responsibility of the
Cambridge Astronomical Survey Unit(http://www.ast.cam.ac.uk/).

A WFS International Review Panel approved three main programmes in the
first year, allocating five--six `dark/grey' weeks per semester to the
WFS. In June 1999 a first year review carried out by PATT and the
International Review Panel confirmed the continuation of the
first year WFS programmes into 2000.

\section{The INT Wide Field Camera}

\begin{figure}
\centerline{
\hspace{0.0cm}
\psfig{figure=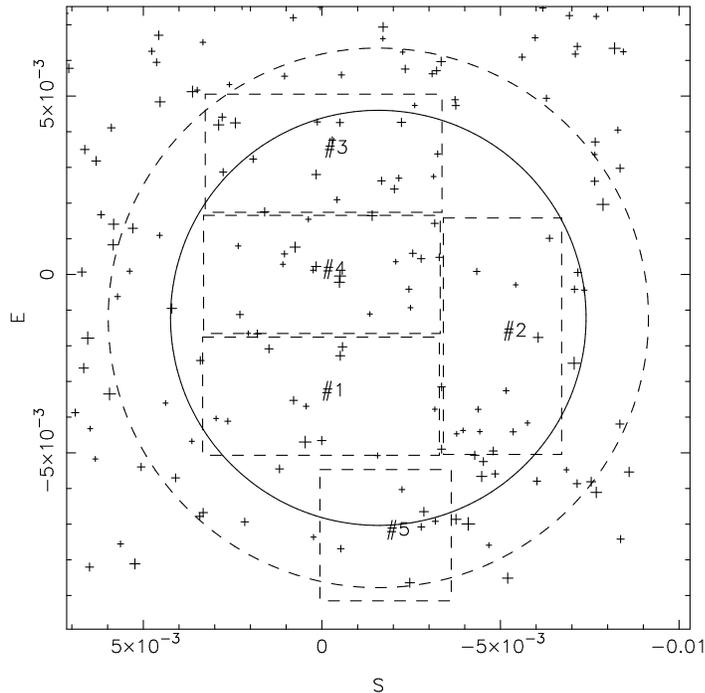,width=14.0cm,angle=-90}
}
\caption
{Layout and normal orientation for the INT Wide Field Camera. The inner
solid circle shows the unvignetted central diameter(40arcmin). The 
dashed circle shows the 50\% vignetted region(52arcmin)}
\label{fig_wfc_geometry}
\vspace{1.0cm}
\end{figure}

The INT Wide Field camera (Ives,Tulloch \& Churchill 1996, see also paper
in these proceedings) is mounted
at the prime focus(f/3) of the 2.5m Isaac Newton telescope on La Palma,
Canary Islands. The camera consists of a close packed mosaic
of 4 thinned EEV42 2kx4k CCDs. The layout is shown in
Figure~\ref{fig_wfc_geometry}.
The CCDs have a pixel size of 13.5 microns corresponding to 0.33 ``/pixel.
The edge to edge limit of the mosaic neglecting the  $\sim$1' inter-chip
spacing is 34.2'. In normal survey mode we use a step size in RA and Dec
of 30' and 20' respectively. This provides $\sim$10\% overlap on all
edges and means that the partially vignetted chip is overlapped completely
to aid photometric calibration.

\section{The current WFS Programmes}

\begin{table}
\caption
[Survey limits]{Nominal photometric limits in 1 arcsec seeing for INT Wide Angle Survey}
\label{table_was_nominal_limits}
\begin{tabular}{|c|c|c|c|c|c|}
\hline
\multicolumn{1}{|c|}{} 
&\multicolumn{1}{|c|}{Exposure} 
&\multicolumn{2}{|c|}{5$\sigma$ detection limit in 1''} 
&\multicolumn{2}{|c|}{1$\sigma$ Surface Brightness limit} 
\\
\multicolumn{1}{|c|}{Waveband}  
&\multicolumn{1}{|c|}{Time} 
&\multicolumn{2}{|c|}{seeing with PSF}
&\multicolumn{2}{|c|}{per square arcsec}
\\
\multicolumn{1}{|c|}{} 
&\multicolumn{1}{|c|}{(secs)}
&\multicolumn{2}{|c|}{ profile fitting} 
&\multicolumn{2}{|c|}{}  
\\
\hline
\multicolumn{1}{|c|}{} 
&\multicolumn{1}{|c|}{} 
&\multicolumn{1}{|c|}{(Vega mag)} 
&\multicolumn{1}{|c|}{(AB mag)} 
&\multicolumn{1}{|c|}{(Vega mag)} 
&\multicolumn{1}{|c|}{(AB mag)} 
\\
\hline
     u          &600   &23.6  &24.5      &25.5     &27.0   \\
     g          &600   &25.2  &25.1      &27.4     &27.3  \\
     r          &600   &24.5  &24.7      &26.7     &26.9  \\
     i          &600   &23.7  &24.1      &25.9     &26.3  \\
     z          &600   &21.7  &22.6      &23.9     &24.8  \\
\hline 
\end{tabular}
\end{table}

\begin{figure}
\centerline{
\hspace{0.0cm}
\psfig{figure=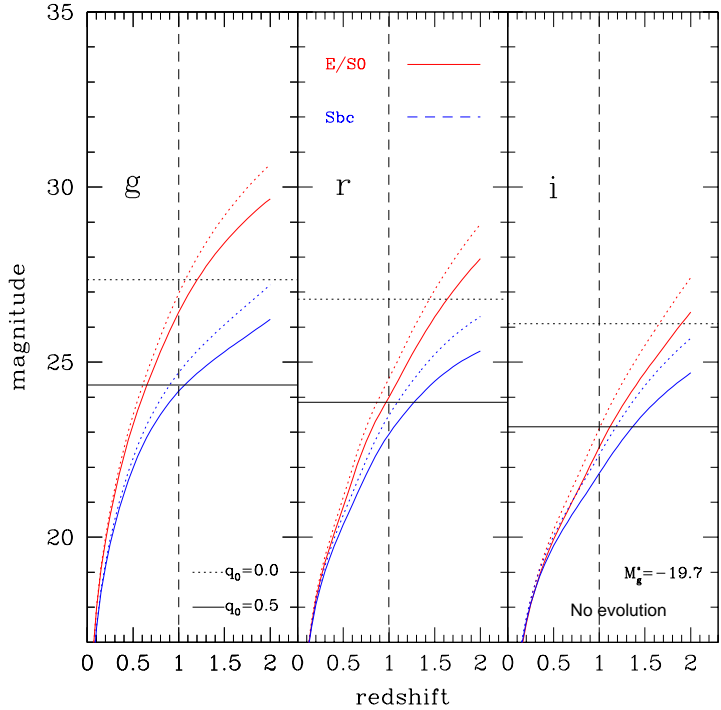,width=10cm,angle=0}
}
\label{fig_was_gals_mags_gri}
\caption
{Observed magnitude versus redshift for L$^*$ 
unevolving galaxy spectral energy distributions. 
The horizontal lines indicate
the magnitude limits(5-$\sigma$ in a 2 arcsec diameter aperture) 
and the surface brightness limits for the proposed INT observations.
The vertical line is drawn at z=1. The upper curves are for an Elliptical 
galaxy SED and the lower pair are for a starforming Spiral galaxy SED.}
\label{fig_int_ao_gal_mags}
\vspace{1.0cm}
\end{figure}

\begin{table}
\caption[Proposed Science Program]{Science Themes(n=23)}
\label{table_was_science_themes}
\vskip 1mm
\begin{tabular}{|l||cccc|r|}
\hline
\hline
Science Theme
&{sample}   
&{surface} 
&{area} 
&{bands}
&{nights\ddag}
\\
{}
&\multicolumn{1}{c}{size}
&\multicolumn{1}{c}{density}
&\multicolumn{1}{c}{deg$^{2}$}
&&\\
{}
&\multicolumn{1}{c}{}
&\multicolumn{1}{c}{deg$^{-2}$}
&&&\\
\hline
Outer Solar System                     &20   & $<$1 &40  &gr     &33 \\
Low mass stars &&&&&\\
\phantom{test} Brown Dwarfs(Pleiades)  &10   &-     &10   &iz    &5  \\
\phantom{test} Young Field BDs         &10    &-    &100  &iz    &27 \\
Young White Dwarfs(Pleiades)           &1     &-    &10    &ugr   &5  \\
Proper Motion study (cool white dwarfs)  &10    &-     &20 &gi   &10 \\
Galactic Structure                     &10$^6$ &10$^4$ &100 &ugriz &77 \\
Stellar Variability  &&&&& \\
\phantom{test} Pop I(CVs, LMXBs)   &20   &-      &20   &ugriz     &16     \\
\phantom{test} Pop II(RR Lyraes)   &100   &-     &20   &gi        &16    \\
Low Surface Brightness Galaxies        &100   &-    &100  &gi    &34 \\
Extremely Red Galaxies                 &100   &-    &100  &gi    &34 \\
2DF Galaxy Survey                      &10$^4$ &5000 &20   &ugriz &16  \\
2DF QSO    Survey                      &2000  &100  &20   &ugriz &16  \\
Supernovae (0.15$<$z$<$0.25)           &20    &-  &20   &gi    &20  \\
Clusters of Galaxies (0.50$<$z$<$1.00) &&&&& \\ 
\phantom{test} Evolution               &100   &5    &20   &gi    &7   \\
\phantom{test} Large Scale Structure   &500   &5    &100  &gi    &34   \\
AGN variability        &1000  &50   &20   &gi     &16  \\
Radio Sources  &&&&&\\
\phantom{test} Large Scale Structure   &5000  &100 &50  &ugi   &25 \\
\phantom{test} High redshift galaxies  &50    &-   &50  &ugi   &25  \\
\phantom{test} Radio Loud Quasars      &500   &10  &50  &ugi   &25  \\
Optically Selected Quasars &&&&&  \\
\phantom{test}(z$<$2)                  &15000  &150  &100 &ug   &34 \\
\phantom{test}(2$<$z$<$5)              &1000   &10   &100 &ugri &67  \\
\phantom{test}(z$>$5)                  &$>$10  &$<$0.5 &100 &riz &43  \\
\hline
Total time for independent programs &    &         &  &    &553\\
\hline
Total time for merged program &   &     &100 & &102  \\
\hline
\end{tabular}
\footnotesize
\noindent {\bf Notes to table:} \newline
\dag assumes an effective area of 0.25deg$^2$; nominal exposures
of 600s in (ugri); 300s in z; overheads per
exposure of 120secs, unless stated otherwise. \newline
\ddag assumes an average night of 8hrs and overall observing
efficiency of 60\% allowing for weather, standards
\end{table}

\begin{table}
\centering
\small
\caption
[Survey Fields]{Nominal Major Survey Regions for INT Wide Angle Survey}
\label{table_was_major_survey_regions}
\vskip 1mm
\begin{tabular}{|l|ccccc|}
\hline
Field  
&{Coordinates} 
&{l,b} 
& size$\dagger$
&bands$\ddagger$
&multi-
\\
& J2000 
& 
&$^\circ$ 
& 
& epoch \\
\hline
NGC Equatorial Strip & 10 $<$ $\alpha$ $<$15 &b$>$$+$30 &75x0.5 &ugriz  &g/r\\
SGC Equatorial Strip & 22 $<$ $\alpha$ $<$03  &b$<$$-$30 &75x0.5 &ugriz  &g/r \\
\hline
WFSJ0220$-$05      & 02~20.0~$-$05 00 &169,$-$53 &3x3 &ugriz  & g/r \\
WFSJ0354+00(SA95)& 03~54.0~+00 00 &187,$-$41 &3x3 &ugriz  &    \\ 
Pleiades     & 03~47.0~+24 00 &167,$-$23 &3x3 &iz     &z   \\
WFSJ0801+40    & 08~01.7~+40 19 &180,$+$30 &3x3 &ugriz  &  \\
Virgo        & 12~39.0~+12 27 &294,$+75$ &3x3 &ugriz  &    \\
WFSJ1610+54      & 16~10.0~+54 30 &84,$+$45  &3x3 &ugriz  &    \\
WFSJ2240+00(SA114)      & 22~40.0~+00 00 &69,$-$49  &3x3 &ugriz  &g/r  \\
\hline
\end{tabular}
Notes: $\dagger$ The actual areal coverage per field is determined by 
the observing time available, observing schedule and observing conditions. 
$\ddagger$ The actual filters may be different eg KPNO B or Harris V etc.
\end{table}

\begin{table}
\centering
\small
\caption
[Survey Fields]{Supplemental survey fields from WAS and the other WFS programs}
\label{table_was_supplemental_survey_regions}
\vskip 1mm
\begin{tabular}{|l|ccccc|l|}
\hline
Field  
&{Coordinates} 
&{l,b} 
& size
&bands$\ddagger$ 
&multi-
&notes
\\
& J2000 
& 
&$^\circ$ 
& 
& epoch 
&\\
\hline
WFSJ0015$+$35  & 03~05~$-$09~35    &115,$-$27    &1.5x1.5 &ugriz  & &Dalton \\
M31            & 00~43~$+$41~17   &121, $-$22  &2@0.5x0.5 &  & &\\
WFSJ0230$+$15  & 02~30~$-$15~30   &155,$+$42    &1.0x1.0 &ugriz &v,i &van Paradijs\\
WFSJ0305$-$09  & 03~05~$-$09~35   &190,$-$54  &0.5x0.5 &ugriz  &  &\\
WFSJ0750$+$20  & 07~50~$+$20~30   &200,$+$22 &1.0x1.0   & &v,i &van Paradijs \\
WFSJ0912$+$41  & 09~12~$+$41~00   &181,$+$43 &1.0x1.0   &ugriz &  & Dalton\\
WFSJ1251$+$27  & 12~51~$+$27~07   &0,$+$90 &1.5x1.5   &ugriz &v,i &van Paradijs \\
WFSJ1610$+$00  & 16~10~$+$00~40    &12,$+$36  &0.5x0.5 & & &SDSS sampler\\
WFSJ1624$+$26  & 16~24~$+$26~34    &45,$+$43  &1.0x1.0  & &v,i &van Paradijs\\
WFSJ1635$+$46  & 16~35~$+$46~30    &72,$+$42   &1.0x1.0 &ugriz  & & Dalton\\
WFSJ1637$+$41  & 16~37~$+$41~16    &65,$+$42   &1.0x1.0 &ugriz  & & \\
WFSJ1720$+$27  & 17~20~$+$27~00    &50,$+$31   &1.0x1.0 &ugriz &v,i &van Paradijs\\
WFSJ2000$+$54  & 20~00~$+$54~57    &89, $+$13  &0.5x0.5 &ugriz & & \\
WFSJ2056$-$04  & 20~56~$-$04~37    &44, $-$27 &1.0x1.0  &ugriz & &\\
WFSJ2345$+$27  & 23~45~$+$27~30    &105,$-$33 &1.0x1.0  &ugriz &v,i &van Paradijs\\
\hline
\end{tabular}
Notes: The exposure times per band vary between these field see the
WFS WWW page for further details. Also, in some fields the WAS program
has added wavebands to those obtained by the original PIs.
$\ddagger$ The actual filters may be different eg KPNO B or Harris V etc.
\end{table}

The main science programmes chosen include a `wide shallow' programme,
a smaller deep area programme, and a programme to address temporal 
variability.
The specific programs are described briefly below:

\subsection{The INT Wide Angle Survey (WAS): co-PI's, McMahon, Irwin, Walton}

This is the largest approved programme approved and includes
sub-projects ranging from determination of cosmological parameters
(via SN Type Ia) to searches for solar system objects.
The underlying philosophy of the WAS survey is encompassed in 
Table~\ref{table_was_science_themes}
where we summarise the time requirements of over 20 topical scientific
programs.  If all these programs were carried out under the normal
PI's based time allocation procedures the total on-sky time required
is almost 600 nights. However, if the programs are combined they can be
executed in around 100 nights. By merging the requirements of the
various programs we end up with a highly efficient observing strategy.
An
important aspect of the reduced time requirements is that the projects
will also be executed quickly.

The limiting magnitudes and wavebands being used are summarised
in Table~\ref{table_was_nominal_limits}. 
Figure~\ref{fig_int_ao_gal_mags}
shows how these
limits transform onto the observational plane for extragalactic studies.
The main survey region are listed in 
Table~\ref{table_was_major_survey_regions}. A number of smaller regions
are also being surveyed as determined by calibration requirements
and the observing 
schedule. In addition, we are adding bands to other programs so that
we can increase the areal coverage of multi-colour data at low cost.
These fields are listed in Table~\ref{table_was_supplemental_survey_regions}.

The WAS program
is the umbrella programme for the WFS project and leads the coordination
efforts  with the other programmes on, for instance, field and filter 
selection, 
to maximise scientific return of the WFS project. All programs remain
autonomous during this procedure so that the peer reviewed science goals
are protected.

Some of the science goals of the WAS are outlined below:

\noindent$\bullet$ {\bf Galactic Studies:} including both halo and disk 
white dwarf 
luminosity function which are relevant to both DM models and to independent 
calibrations of the 
Hubble time; stellar density distributions towards the NGP, to improve extant 
K$_z$ determinations of the local DM; stellar counts towards the anti-centre 
and other widely spaced directions, to determine the stellar warp and refine
models of Galactic structure

\noindent$\bullet$ {\bf Clusters of Galaxies:} 
the aim is to determine the space density 
and cluster-cluster correlation function over the range 0.5$<$z$<$1.0.
Galaxy clusters are the largest gravitationally-bound structures in
the Universe, and the study of their abundance and evolutionary
history with look-back-time places strong constraints on cosmological
parameters and the primordial power spectrum that gave rise to the
observed large scale structure.

\noindent$\bullet$ {\bf Radio Sources \& Radio Galaxies:}  
Deep optical identification of radio sources allows: accurate counts of 
different types of host along mJy tracks in the P-z plane, studies
of radio source 
luminosity evolution; multi-band investigation of giant-E standard candles;  
the largest known sample of low-luminosity RGs with good photometry;
large-scale structure from photometric redshifts and cell counts in 
redshift slices; accurate optical positions of FIRST sources for 
WYFFOS/2DF followup.

\noindent$\bullet$ {\bf Intermediate redshift Type 1a Supernovae:} Whilst
dramatic progress has been made in the determination of 
the fundamental cosmological parameters ($\Omega$, $\Lambda$)
in the last two years, the analysis is now limited by 
systematic errors. Identifying
$\sim$20 Type 1a Supernovae in the critical range 0.1$<$z$<$0.4
will allow a detailed treatment of these systematic errors.

The WAS also incorporates two independent distinct science programmes in 
the spring
semester centred on Virgo and the North Galactic Pole. In fact, in
the proposal submission procedure many co-I's of the WAS program submitted
discrete proposals.

\begin{itemize}
\item {\bf A multicolour survey of the Virgo Cluster: PI, Davies}
This aims to obtain the galaxy luminosity function (LF) of the
Virgo  cluster as a function of colour and position in the
cluster.

\item{\bf The Millennium Galaxy Catalogue (MGC): PI, Driver}
The MGC will provide a complete and local galaxy catalogue. This
survey is being carried out in the B band and lies in a region of
sky covered by the 2DF redshift survey.
\end{itemize}

\subsection{A Deep UBVRI Imaging Survey with the WFC: PI, Dalton}

This programme is carrying out deep imaging
of 10 deg$^2$ to a limiting magnitude of B=26 and I=24.5. It will enable
the study of the evolution of galaxy clustering as a function of
colour at faint magnitudes and provide a catalogue of rich galaxy
clusters at intermediate red shifts.

\subsection{Faint Sky Variability Survey (FSVS): PI, van Paradijs}

This programme is searching an area of $\sim$10 deg$^2$, studying photometric
and astrometric variability on scales of one hour to a year to a
magnitude of V=25. Example areas of investigation include: the
evolution of specific galactic populations (e.g. CV's, RR Lyraes, halo
AGB stars, brown \& white dwarfs, Kuiper-Edgeworth belt objects, 
sdB stars), the
structure of the galactic halo, statistics of optical transients
related to $\gamma$-ray bursts, and  deep proper motion studies. 


\section{Choice of survey regions and photometric bands}
In order to maximise the scientific value of the WFS data the
WAS survey is concentrating on fields that are equatorial and hence
follow-up can be carried out from telescopes on both hemispheres. This
simple consideration doubles the scientific return of the survey. We also
deliberately centred some of the fields on Landolt photometric calibration
fields ie SA95 and SA114.

The choice of photometric wavebands was relatively straight forward. We decided
to use bands similar to the SDSS bands(Fukugita etal, 1996). Note
our u and z bands are not identical to the SDSS bands. See the WFS
WWW pages for further details. The choice of the SDSS bands means that
the INT surveys will be directly comparable with work carried out as
part of the SDSS. Interestingly, the SDSS g band is very close to the
UKST $B_J$ band. However, manufacturing delays have meant that we had
to start the survey using the standard INT filter set. 

\section{Survey Coverage to Date}
Survey data is being obtained on a monthly basis and thus a summary of
the data obtained will soon be out of date. A complete summary of
observations obtained is kept on-line at
http://www.ast.cam.ac.uk/\~{}wfcsur/status.

The situation at the end of May 1999 was that $\sim$60 deg$^2$ had been
observed in the  first ten months of the survey.

\section{Data Products}

The data products currently available for access include:
\begin{itemize}
\item Observing logs built from the FITS headers
\item A SYBASE WWW user interface to access the raw and processed data
\item Library bias frames, flatfield frames, defringing frames and
    non-linearity corrections
\item Colour equations for all filters
\item Processed 2D image maps, with a full record of processing steps in
  the FITS headers
\item Astrometric calibration, with the World Coordinate System in 
the FITS headers
\item Photometric calibration --- zero points and extinction 
\end{itemize}

In the coming months the data products provided will be expanded after
some quality control to include:
\begin{itemize}
\item Object catalogues, generated using APM based routines (Irwin, 1985)
and SExtractor (Bertin \& Arnouts 1996).
\end{itemize}


\section{Further Information}
Further information about the INT Wide Field Imaging Survey can
be obtained at the Isaac Newton Groups WWW page (www.ing.iac.es/WFS) 
or the UK mirror(www.ast.cam.ac.uk/ING/WFS). In addition, the Wide
Angle Survey has as a WWW page at www.ast.cam.ac.uk/\~{}rgm/int\_sur/.
Further details of the pipeline processing are contained in a paper by
Irwin and Lewis(these proceedings).

\paragraph*{Acknowledgements} The authors would like to acknowledge
the unsung builders of the INT Wide Field Camera, the various peer review
committee members who have guided the project since it inception in 1991
and the encouragement of
their many colleagues during the long gestation period of the INT Wide Field
Survey project.

{}

\end{document}